\documentclass[iicol]{sn-jnl}

\usepackage{multirow}%
\usepackage{amsmath,amssymb,amsfonts}%
\usepackage{mathrsfs}%
\usepackage[title]{appendix}%
\usepackage{xcolor}%
\usepackage{textcomp}%
\usepackage{manyfoot}%
\usepackage{booktabs}%
\usepackage{algorithm}%
\usepackage{algorithmicx}%
\usepackage{algpseudocode}%
\usepackage{listings}%
\usepackage{comment}
\usepackage{url}
\usepackage{mathtools}
\usepackage{empheq}

\DeclareMathOperator{\eqEOM}{\overset{\textrm{ \tiny EOM}}{=}}
\DeclareMathOperator{\Christo}{\Gamma}

\begin{document}

\title[]{A Hamiltonian approach to the gradient-flow equations in information geometry}


\author*[1]{\fnm{Tatsuaki} \sur{Wada}}\email{tatsuaki.wada.to@vc.ibaraki.ac.jp}

\author[2]{\fnm{Antonio M.} \sur{Scarfone}}\email{antonio.scarfone@polito.it}

\affil*[1]{\orgdiv{Region of Electrical and Electronic Systems Engineering}, \orgname{Ibaraki University}, \orgaddress{\street{Nakanarusawa-cho}, \city{Hitachi}, \postcode{316-8511}, 
 \country{Japan}}}

\affil[2]{\orgname{Istituto dei Sistemi Complessi, Consiglio Nazionale delle Ricerche (ISC-CNR), c/o Politecnico di Torino}, \orgaddress{\street{Corso Duca degli Abruzzi 24}, \city{Torino}, \postcode{I-10129}, 
 \country{Italy}}}


\abstract{We have studied the gradient-flow equations in information geometry from a point-particle perspective.
Based on the motion of a null (or light-like) 
particle in a curved space, we have rederived the Hamiltonians
which describe the gradient-flows in information geometry.}




\maketitle

\section{Introduction}
\label{intro}

Information geometry (IG) \cite{Amari} is a useful and powerful framework for studying some families of probability distributions by identifying the space of probability distributions with a differentiable manifold endowed with Fisher metric as a Riemann metric and $\alpha$-connection as an affine connection. In IG,  the Riemann metric is obtained from the Hessian of a potential function, and the fluctuations play important role. Especially the so-called fluctuation-response relations are related with the Hessian metric \cite{WS15}.
On the other hand, the gradient-flow equations are useful for some optimization problems. The gradient flows on a Riemann manifold follow the direction of gradient descent (or ascent) in the landscape of a potential functional, with respect to the curved structure of the underlying metric space. The IG studies on the gradient systems were originally performed independently by Nakamura~\cite{N94} and Fujiwara-Amari~\cite{FA95}. A remarkable feature of their works is that a certain kind of gradient flow on a dually flat space can be expressed as a Hamilton flow. Later, several works on this issue have been done from the different perspectives. Malag\'o and Pistone \cite{MP15} studied the natural gradient flows in the mixture geometry of a discrete exponential family.
Boumuki and Noda \cite{BN16} studied the relationship between the Hamiltonian-flows and gradient-flows from the perspective of symplectic geometries. 
Chirico et~al. \cite{CMP22}  provided an information geometric formulation of classical mechanics on the statistical manifold.
Furthermore Pistone \cite{Pistone} studied Lagrangian function on the finite state space statistical bundle.
Together with our collaborators, we also studied the same issue by some heuristic approaches and related to some different fields. In Ref.~\cite{WSM21},  we studied the gradient-flow equations based on the generalized eikonal equation for a simple thermodynamic system and introduced a mock time evolution in IG as a Hamilton-Jacobi dynamics. We studied \cite{WSM23} the same issue from the perspective of geometric optics and related the gradient-flows in IG to the light trajectories in anisotropic optical media by using Huygens' equations. Furthermore the analytical mechanical properties concerning the gradient-flow equations in IG are studied \cite{CW24} and discussed the deformations of the gradient-flow equations which lead to Randers-Finsler metrics \cite{Randers}. Ref. \cite{W23} provided a Weyl geometric approach. 
Through these studies we realize the importance of treating space and time on equal footing, which is an essence of Einstein's relativity \cite{Blau}. In addition, it is known in general relativity that in a suitable coordinate system, the physical equations have simple forms and clear physical meanings \cite{G18}.  

Through our previous studies \cite{WSM21,WSM23, CW24}, we already related  the gradient-flows in IG 
to the Hamilton-flows based on the physical concepts such as a light-ray, refractive index in the geometrical optics,  in which optical (or Fermat) metrics
play a central role. It is noted \cite{F13} that based on Fermat's principle, the spatial part of null geodesic in $(N+1)$-dimensional spacetime is regarded
as the geodesic of the corresponding $N$-dimensional optical geometry.  A null geodesic equation in pseudo-Riemann space can be considered as a Hamilton-Jacobi equation, and
the complete integrability is a key for solving the Hamilton-Jacobi equations. 
In this contribution we take a different approach based on Hamiltonian systems to the gradient-flows in IG by considering the complete integrability and by analyzing the motion of a null (or light-like) particle in a pseudo-Riemann metric.

The rest of the paper consists as follows.
In Section \ref{IG}, we first review some basics of IG and the associated gradient-flow equations. 
In Section \ref{CompInt} we consider the analytical mechanics based on Cartan's theory \cite{Cartan} on the complete integrability of Pfaffian systems.
We obtain the form of a Hamiltonian which satisfies the complete integrability condition and relate it to the Hamiltonian describing the gradient-flows
in IG.
Section \ref{Null}  shows that the gradient-flows in IG are related to the motions of a light-like particle in a curved space characterized by a pseudo-Riemann metric, which are analyzed in the fields of general relativity.
 The final Section \ref{conclusions} is devoted to our conclusions and perspective.
Appendix \ref{theproof} shows the proof of the condition \eqref{condExactInt} for the complete integrability of Pfaffian equation.
Appendix \ref{conformal} provides the relation between the conformal scaling and reparametrization.
Appendix \ref{Zform} shows the explicit relation of the stationary metric discussed in Section \ref{Null} to the Zermelo form.

We would like to emphasize that our approaches to IG are different from the conventional method as follows.
In conventional method \cite{Amari} of IG, the natural ($\theta$- or $\eta$-) coordinate space (or dually-flat manifold) is characterized with Fisher metric $g$ and the $\alpha$-connections, which provide the parallel translation rule. In addition, unlike Riemann geometry, the metric $g$ is only used to determine the orthogonality but not used to determine a distance in the natural coordinate spaces. The $\theta$- and $\eta$-coordinate systems are regarded as the dual coordinates on the same manifold. In contrast, in our perspective, the $\theta$- and $\eta$-coordinate spaces are regarded as the two different spaces (or basis manifolds) in general. When the $\theta$-space belongs to
a curved space which is regarded as a basis manifold $\mathcal{M}$, the corresponding $\eta$-space belongs to the cotangent space $T^{\star} \mathcal{M}$. On the other side,
 when the $\eta$-space belongs to a curved space $\mathcal{N}$, the corresponding $\theta$-space belongs to the cotangent space $T^{\star} \mathcal{N}$. 


Throughout the paper, we use Einstein's summation convention
and assuming that a Latin index (e.g., $i, j, k, \dots$) runs from $1$ to $N$, while a Greek index (e.g., $\mu, \nu, \dots$) runs from $0$ to $N$.

\section{Information Geometry and Gradient-Flow Equations}
\label{IG}
Here, some basics of IG and the gradient-flow equations are reviewed.

\subsection{Information Geometry}
In IG \cite{Amari}, the so-called \textit{dually-flat structures} are important. 
For a given set of some functions $F_i(x), i=1, \dots, N$,
 the $\theta$-parametrized probability distribution function (pdf)
\begin{align}
  p_{\theta}(x) = \exp \left[ \theta^i  F_i(x) - \Psi(\theta)  \right],
  \label{exp-pdf}
\end{align}
is called an exponential pdf. Here $\Psi(\theta)$ is determined from the normalization of $p_{\theta}(x)$ as $\Psi(\theta) = \ln \left[ \int dx \exp (\theta^i F_i(x) ) \right]$.
A manifold of probability distribution, which is called \textit{statistical manifold} $(\mathcal{ M}, g, \nabla, \nabla^\star)$, is characterized
by a pseudo-Riemannian metric $g$, and torsion-less dual affine connections
$\nabla$ and $\nabla^\star$. 
For a given convex function $\Psi(\theta)$ together with its dual convex function $\Psi^\star(\eta)$,
one can construct the dually-flat structure as follows.
From the dual convex functions $\Psi^\star (\eta)$ and $\Psi (\theta)$,
the associated dual affine coordinates $\theta^i$ and $\eta_i$ 
are obtained as
\begin{align} 
 \theta^i = \frac{\partial \Psi^\star(\eta)}{\partial \eta_i}, \quad 
\eta_i = \frac{\partial \Psi(\theta)}{\partial \theta^i},
\label{theta-eta} 
\end{align}
respectively. These convex functions are Legendre dual to each other
\begin{align}
    \Psi^\star(\eta) = \theta^i \, \eta_i - \Psi(\theta).
    \label{LegendreT}
\end{align}

Taking logarithm of both sides of \eqref{exp-pdf} and taking expectation, we have
\begin{align}
 {\rm E}_{p_{\theta}} [ \ln  p_{\theta}(x) ] = \theta^i   \, {\rm E}_{p_{\theta}} [ F_i(x) ] - \Psi(\theta),
 \label{exprel}
\end{align}
where
\begin{align}
 {\rm E}_{p_{\theta}} [ f(x) ] := \int dx \, p_{\theta}(x) f(x),
 \end{align}
 denotes the expectation value of a function $f(x)$ with respect to $p_{\theta}(x)$.
Comparing  \eqref{LegendreT} to \eqref{exprel}, we see that
\begin{align}
  \eta_i = {\rm E}_{p_{\theta}} [ F_i(x) ],
\end{align}
and
\begin{align}
  \Psi^\star(\eta) = {\rm E}_{p_{\theta}} [ \ln  p_{\theta}(x) ] =: - S(\eta),
  \label{-S}
\end{align}
where $S(\eta)$ is the entropy.
The positive definite matrices $g_{ij}(\theta)$ and $g^{ij}(\eta)$ are obtained from the Hessian matrices
of the convex function $\Psi(\theta)$ and $\Psi^\star(\eta)$ as 
\begin{align} 
g_{ij} ( \theta)& = 
\frac{\partial \eta_i}{\partial \theta^j} = 
\frac{\partial^2 \Psi( \theta)}{\partial \theta^i \partial \theta^j}, \nonumber \\
g^{ij} ( \eta) &= 
\frac{\partial \theta^i}{\partial \eta_j} = 
\frac{\partial^2 \Psi^\star(\eta)}{\partial \eta_i \partial \eta_j},
\label{g} 
\end{align}
respectively.
These matrices satisfy the relation $g^{ij} (\eta) \,
g_{jk} (\theta) = 
\delta^i_k$,
where $\delta^i_k$ denotes Kronecker's delta.

Since connection coefficients $\Christo$ are not tensors, there exists a coordinate system in which all connection coefficients become zero and such
a coordinate system is called an \textit{affine coordinate}.
The $\alpha$-connection $\nabla^{(\alpha)}$ \cite{Amari}, which is a one-parameter extension $ \left\{ \nabla^{\alpha} \right\}_{\alpha \in \mathbb{R}}$ of Levi-Civita's connection $\nabla^{(0)}$, and its dual $\nabla^{\star (\alpha)}$ are defined by their coefficients as
\begin{align}
   \Christo^{(\alpha)}{}_{ijk}(\theta) &:= \frac{(1 - \alpha)}{2} C_{ijk}(\theta), \nonumber \\
   \Christo^{\star (\alpha) \; ijk} (\eta) &:= \frac{(1 + \alpha)}{2} C^{ijk}(\eta),
   \label{aGamma}
\end{align}
respectively.
Here $C_{ijk}(\theta)$ and $C^{ijk}(\eta)$ are the total symmetric cubic tensors (\textit{Amari-Chentsov} tensors)
\begin{align}
   C_{ijk}(\theta)  &:= \frac{\partial^3 \Psi(\theta)}{\partial \theta^i \partial \theta^j \partial \theta^k}, \nonumber \\
   C^{ijk}(\eta)  &:= \frac{\partial^3 \Psi^\star(\eta)}{\partial \eta_i \partial \eta_j \partial \eta_k}.
\end{align}
Among the $\alpha$-connections, $\alpha = \pm 1$ play a central role \cite{Amari}. One readily sees, from \eqref{aGamma}, that the connection coefficients $\Christo^{(1)}{}_{ijk}(\theta)$ of $\nabla^{(1)}$ ($\Christo^{\star (-1) \; ijk} (\eta)$ of $\nabla^{\star (-1)} $) vanish and hence the $\theta$-coordinates ($\eta$-coordinates) are affine for the connection $\nabla^{(1)}$ ($\nabla^{\star (-1)} $) .

A divergence $D(p, q)$ of a set of two states $p$ and $q$ is a non-negative function providing a measure how much they differ.
Some known examples of the divergences are relative entropy (or Kullback-Leibler divergence) and $f$-divergence.
In IG the $\theta$- and $\eta$-divergence functions are
\begin{subequations}
\begin{align}
  D(\theta, \theta_{\rm r}) &:=   \Psi(\theta) - \Psi(\theta_{\rm r}) - \eta_i^{\rm r} ( \theta^i -\theta^i_{\rm r} ),  \label{thetaD} \\
  D(\eta, \eta^{\rm r}) &:=   \Psi^{\star}(\eta) - \Psi^{\star}(\eta_{\rm r}) - \theta^i_{\rm r} ( \eta_i -\eta_i^{\rm r} ),
\end{align}
\end{subequations}
respectively. Here the $\theta_{\rm r}$ (or $\eta^{\rm r}$) denotes the $\theta$- (or $\eta$-) vector of a reference state.
When $\theta = \theta_{\rm r}$, the $\theta$-divergence $D(\theta, \theta_{\rm r})$ vanishes and similarly the $\eta$-divergence $D(\eta, \eta^{\rm r}) $ vanishes when $\eta = \eta^{\rm r}$.

\smallskip 

\subsection{Gradient-Flow Equations}
The gradient-flow equations \cite{N94,FA95,BN16} in IG are briefly explained here.
The gradient-flow equations with respect to the $\theta$-divergence  function $D(\theta, \theta_{\rm r})$ with a given fixed $\theta_{\rm r}$ are 
\begin{align} 
\frac{d \theta^i}{d t} &=  g^{ij} ( \theta) \, \frac{\partial D(\theta, \theta_{\rm r})}{\partial \theta^j}  \nonumber \\
&= g^{ij} ( \theta) \, \left( \frac{\partial \Psi(\theta)}{\partial \theta^j} - \frac{\partial \Psi(\theta)}{\partial \theta^j} \Big\vert_{\theta^j = \theta_{\rm r}^j} \right),
\label{theta-gradEq}
\end{align}
in the $\theta$-coordinate system.
By using the properties \eqref{theta-eta} and \eqref{g}, the left-hand side of \eqref{theta-gradEq} is rewritten by
\begin{align}
\frac{d \theta^i}{d t} &= 
\frac{\partial \theta^i}{\partial \eta_j} 
\frac{d \eta_j}{d t} = 
g^{ij} ( \theta) 
\frac{d \eta_j}{d t},
\end{align}
and applying \eqref{theta-eta}  to the right-hand side of \eqref{theta-gradEq} leads to $g^{ij} ( \theta) \, (\eta_j - \eta^{\rm r}_j )$.
Consequently, the gradient-flow equations \eqref{theta-gradEq} in the $\theta$-coordinate system are equivalent
to the linear differential equations
\begin{align}
\frac{d \eta_i (t)}{d t} = 
\eta_i (t) - \eta^{\rm r}_i,
\end{align}
in the $\eta$-coordinate system. This linearization is one of the merits due to the dually-flat structure \cite{Amari} in IG.

The other set of gradient-flow equations are given by
\begin{align} 
\frac{d \eta_i}{d t} &= - g_{ij} ( \eta) \frac{\partial D(\eta, \eta^{\rm r})}{\partial \eta_j} \nonumber \\
&= -g_{ij} ( \eta) \! \left( \frac{\partial \Psi^{\star}(\eta)}{\partial \eta_j} - \frac{\partial \Psi^{\star}(\eta)}{\partial \eta_j} \Big\vert_{\eta_j=\eta^{\rm r}_j }\right),
\label{eta-gradEq}
\end{align}
in the $\eta$-coordinate system.
Similarly, they are equivalent to the linear differential equations
\begin{align}
\frac{d \theta^i (t)}{d t} = 
- \theta^i (t) + \theta_{\rm r}^j,
\label{linearDE}
\end{align}
in the $\theta$-coordinate system.
In the previous works \cite{WSM21,WSM23,CW24,W23}, the gradient-flow equations with respect to the $\theta$- or $\eta$-potential functions were considered. These cases correspond to $\theta_{\rm r}^i = \eta^{\rm r}_i = 0$, since, for example, the gradients of the $\theta$-divergence \eqref{thetaD} in the cases are equal
to the gradients of the $\theta$-potential function. i.e., 
\begin{align}
  \frac{ \partial}{\partial \theta^i} D(\theta, \theta_{\rm r}) \Big\vert_{ \theta_{\rm r}^i = \eta^{\rm r}_i = 0} = \frac{ \partial \Psi(\theta) }{ \partial \theta^i}.
 \end{align}

It is worth emphasizing that the two sets of differential equations \eqref{theta-gradEq} and \eqref{eta-gradEq}
describe different processes in general \cite{WSM23, CW24}.
In addition, the evolutional parameter $t$ in the gradient-flow equations \eqref{theta-gradEq} and \eqref{eta-gradEq} is a non-affine parameter.
Recall that  a parameter $s$ is affine if the geodesic equations of a curve $x^i = x^i(s)$ are in the form:
\begin{align}
  \frac{d^2 x^i(s)}{d s^2} + \Christo^{i}{}_{jk} (x) \frac{d x^j(s)}{ds}   \frac{d x^j(s)}{ds}  = 0.
\end{align} 
For example, we see from \eqref{linearDE} that
\begin{align}
\frac{d^2 \theta^i}{d t^2} = - \frac{d \theta^i}{d t}.
\end{align}
This is the non-affinely parametrized geodesic (or pre-geodesic) equations in the $\theta$-space, in which the $\theta$-coordinates are of course affine and $\Christo^{i}{}_{jk} (\theta)=0$.

As we mentioned in Introduction,  Ref. \cite{WSM23, CW24}  have related the gradient-flows \eqref{theta-gradEq}  and \eqref{eta-gradEq} in the case of $\theta_{\rm r}^i = \eta^{\rm r}_i = 0$ to the Hamilton-flows characterized by the Hamiltonians
\begin{subequations}{\label{OurH}}
\begin{align}
 H(\theta, \eta) &= \sqrt{g^{ij}(\theta) \eta_i \eta_j} -  \sqrt{\eta^2(\theta)}, \nonumber \\  \textrm{with} \;  & \eta^2(\theta) := g^{ij}(\theta) \frac{\partial \Psi(\theta)}{\partial \theta^i} \frac{\partial \Psi(\theta)}{\partial \theta^j},
 \label{H4theta} \\
 H(\eta, -\theta) &= \sqrt{g_{ij}(\eta) \theta^i \theta^j} - \sqrt{\theta^2(\eta)},  \nonumber  \\  \textrm{with} \;   & \theta^2(\eta) := g_{ij}(\eta) \frac{\partial \Psi^{\star}(\eta)}{\partial \eta_i} \frac{\partial \Psi^{\star}(\eta)}{\partial \eta_j},
 \label{H4eta}
 \end{align}
\end{subequations}
or equivalently
\begin{subequations}{\label{Our2ndH}}
\begin{align}
 H(\theta, \eta) &= \frac{1}{2} g^{ij}(\theta) \eta_i \eta_j - \frac{1}{2} \eta^2(\theta), 
 \label{2ndH4theta} \\
 H(\eta, -\theta) &= \frac{1}{2} g_{ij}(\eta) \theta^i \theta^j - \frac{1}{2} \theta^2(\eta), 
 \label{2ndH4eta}
 \end{align}
\end{subequations}
respectively\footnote{The quantity $\theta^2(\eta)$ and $\eta^2(\theta)$ were denoted as $n^2$ and $(n^{\star})^2$, respectively, in our previous study \cite{WSM23}, and  $n$ or $n^{\star}$ was regarded as the refractive index of an optical medium.}.  It must be noted that $\eta^2 (\theta)$ is a function of $\theta^i$ only and $\theta^2(\eta)$ is a function
of $\eta_i$ only.

The associated evolutional parameter is $t$, which is a non-affine parameter.
Note also that $H(\theta, \eta)$ and $H(\eta, -\theta)$ are related through the canonical transformation $(\theta^i, \eta_i)$ to $(\eta_i, -\theta^i)$ \cite{CW24}.
In Section \ref{CompInt} we will show  the complete integrability of Pfaffian systems leads to a Hamiltonian $H(x, p)$ which is homogeneous of first order in the variable $p$.
In Section \ref{Null} we will rederive the above Hamiltonians by considering the motion of a null (or light-like) particle in a curved space.

It is  worth noting that the scalar field $ \eta^2(\theta)$ characterizes
the rate of the $\theta$-potential, since it is related to the $\theta$-potential function $\Psi(\theta)$ as follows.
\begin{align}
  \frac{d \Psi(\theta)}{d t} &= \frac{\partial \Psi(\theta)}{\partial \theta^i} \frac{d \theta^i}{d t}  \nonumber \\
  &= g^{ij}(\theta) \frac{\partial \Psi(\theta)}{\partial \theta^i} \left(  \frac{\partial \Psi(\theta)}{\partial \theta^j}  - \frac{\partial \Psi(\theta)}{\partial \theta^j} \Big\vert_{\theta^j = \theta_{\rm r}^j}  \right) \nonumber \\ 
  &= \eta^2(\theta) - g^{ij}(\theta) \eta_i  \eta^{\rm r}_j .
  \label{dPsi_dt}
\end{align}
where the relations \eqref{theta-eta} and \eqref{theta-gradEq} are used. Similarly, the scalar field $ \theta^2(\eta)$ characterizes
the rate of the $\eta$-potential as  
\begin{align} 
  \frac{d \Psi^{\star}(\eta) }{ dt}  = -\theta^2(\eta) + g_{ij}(\eta) \theta^i \theta_{\rm r}^j.
  \label{dPhistar2dt}
\end{align}
Since  $-\Psi^{\star}(\eta)$ is the entropy $S(\eta)$ in \eqref{-S}, the scalar field $\theta^2(\eta)$ characterizes 
the rate of the entropy $d S(\eta) / dt$ in the gradient-flows.


\subsection{Randers-Finsler deformation of the gradient-flow equations}
Here we briefly review the Randers-Finsler (RF) deformation \cite{CW24} of the gradient-flow equations.

Finsler space is a general space based on the line-element $d \ell = F(x, dx)$, where $F(x, dx)>0$ for $dx \ne 0$ is a function in the tangent space $T_x \mathcal{M}$,
and is a homogeneous function of first order in $dx$.
The function $F(x, dx)$ is called \textit{Finsler function}, which provides the metric tensor
\begin{align}
 g_{ij}(x, dx) = \frac{1}{2} \frac{\partial F^2(x, dx)}{\partial dx^i \partial dx^j},
\end{align}
in the tangent space.
Finsler geometry is a generalization of Riemann geometry with no restriction of the quadratic form $F^2 = g_{ij}(x) dx^i dx^j$.

Randers \cite{Randers} functions were derived from his research on general relativity and have been applied in many fields of sciences. 
Randers function $F(x, dx)$ is a special class of Finsler function and is composed of a Riemannian line-element $ \sqrt{a_{ij}(x) dx^i dx^j}$ and one-form $b_i(x) dx^i$ as
\begin{align}
    F(x, dx) = \sqrt{a_{ij}(x) dx^i dx^j} +  b_i(x) dx^i,
\end{align}
which is homogeneous of first order in $dx^i$. For the arc length $s$ of a curve parameterized by $\tau$ between two points given by \begin{align}
  s=  \int_A^B  F(x, dx) = \int _A^B L_{\rm RF} \left(x, \frac{d x}{d \tau} \right) d \tau,
\end{align}
the corresponding RF Lagrangian is 
\begin{align}
    L_{\rm RF} \! \left(x, \frac{d x}{d \tau} \right) \! = \! \sqrt{a_{ij}(x) \frac{dx^i}{d \tau} \frac{ dx^j}{d \tau}} \! + \!  b_i(x) \frac{dx^i}{d \tau}.
\end{align}

In Ref. \cite{CW24}, based on the Randers functions, the gradient-flow equations with respect to the $\theta$-potential function $\Psi(\theta)$  were deformed as
\begin{align}
  \frac{ d \theta^i}{dt} &= g^{ij}(\theta) \left( \frac{\partial \Psi(\theta)}{\partial \theta^j} - A_j(\theta) \right) \nonumber \\ 
     &= g^{ij}(\theta) \Big( \eta_j - A_j(\theta) \Big),
  \label{RFtheta-gradEq}
\end{align}
where each $A_j(\theta)$ denotes a function of $\theta$ due to this deformation.
It is worth noting  that the  gradient-flow equations  \eqref{theta-gradEq} with respect to the divergence $D(\theta, \theta_r)$ correspond to
the RF deformed equations \eqref{RFtheta-gradEq} in which $A_j(\theta) = \eta^{\rm r}_j$.

Now introducing the quantity $\chi^2(\theta)$ as 
\begin{align}
 \chi^2(\theta) := g_{ij}(\theta) \frac{d \theta^i}{dt}  \frac{d \theta^j}{dt},
 \label{chi2}
 \end{align}
which is the deformation of $\eta^2(\theta)$ in \eqref{H4theta}. Indeed, by utilizing \eqref{RFtheta-gradEq}, we see that 
\begin{align}
  \chi^2(\theta)  = g^{ij}(\theta) \Big(  \eta_i - A_i(\theta) \Big) \Big( \eta_j - A_j(\theta) \Big),
  \label{rel_chi2}
  \end{align}
which reduces to $\eta^2(\theta)$ when $A_j(\theta) \to 0$.
Note that the quantity  $\chi^2(\theta)$ characterizes the ratio
of the infinitesimal arc-length square $ds^2 = g_{ij}(\theta) d\theta^i d\theta^j$ to $dt^2$, i.e.,  
\begin{align}
 \chi^2(\theta) = \frac{ds^2 }{ dt ^2}.
 \end{align} 
 
 Next we introduce $d \tilde{t}$ such as
  \begin{align}
 \chi^2(\theta) =  \frac{d\Psi(\theta)}{d \tilde{t}}.
 \end{align}
 Then, from the RF deformation \eqref{RFtheta-gradEq} we have
  \begin{align}
d \tilde{t}  &=  \frac{1}{\chi^2(\theta)} \, d\Psi(\theta) =  \frac{1}{\chi^2(\theta)} \,  \frac{ \partial \Psi(\theta)}{\partial \theta^i} d \theta^i \nonumber \\
  &=  \frac{g_{ij}(\theta) }{\chi^2(\theta)}\frac{d \theta^j}{dt} d \theta^i +  \frac{A_i(\theta)}{\chi^2(\theta)} \, d \theta^i \nonumber \\
   & =   dt + \frac{A_i(\theta)}{\chi^2(\theta)} \,  d \theta^i.
 \end{align}
Here we used
\begin{align}
  dt = \sqrt{ \frac{g_{ij}(\theta)}{ \chi^2(\theta)} d \theta^i d \theta^j},
 \end{align}
 which is obtained from \eqref{chi2}.
 The Randers function $d \tilde{t}$ reduces to $dt$ when $A_i(\theta) \to 0$.
 The corresponding RF Lagrangian is
 \begin{align}
   \mathcal{L}_{\rm RF} \! \left( \theta, \! \frac{d \theta}{dt} \right) \! = \! \sqrt{ \frac{g_{ij}(\theta)}{ \chi^2(\theta)} \frac{d \theta^i}{dt} \frac{ d \theta^j}{dt}} 
  \! + \! \frac{A_i(\theta)}{\chi^2(\theta)} \,  \frac{d \theta^i}{dt}.
   \label{RFLag}
 \end{align}

\section{Complete integrability and geodesic Hamiltonian}
\label{CompInt}
Here we discuss the complete integrability concerning a certain kind of Hamiltonian in analytical mechanics.
Let us begin with a brief review on the complete integrability of Pfaffian systems by \'{E}lie Cartan \cite{Cartan}.
He extended Poincar\'{e}'s theory on the integral invariant, and showed that the one-form
\begin{align}
 \omega_{\rm PC} := p_j dx^j - H dt, 
 \label{PCform}
\end{align}
is very useful for studying the time evolution in classical mechanics under the action of a Hamiltonian $H=H(x, p, t)$. The one-form  $\omega_{\rm PC}$ is defined in the extended configuration space of $(x, t) \in \mathcal{M} \times \mathbb{R}$, and is known as the Poincar\'e-Cartan one-form \cite{Arnold}.

Now consider the complete integrability of the Pfaffian equation
\begin{align}
       \omega_{\rm PC} = 0,
  \label{Pfaffsys}
\end{align} 
for the Poincar\'e-Cartan one-form $\omega_{\rm PC}$ \eqref{PCform}.
Recall that the Pfaffian system is said to be \textit{completely integrable} if the integral surface of \eqref{Pfaffsys} is given by the equations $S_a = \textrm{constant}$, where $S_a$ is a potential function and $\omega_{\rm PC} = d S_a$. In other words, there exists a differentiable function $S_a = S_a(x, t)$ such that the Pfaffian equation \eqref{Pfaffsys} is equivalent to $dS_a = 0$.
Soon later, we will see that this function $S_a$ is the \textit{action} in analytical mechanics.
According to Frobenius integrability theorem \cite{Cartan}, the necessary and sufficient condition of the complete integrability of \eqref{Pfaffsys}
is $d \omega_{\rm PC} \wedge \omega_{\rm PC} = 0$.
With the help of Hamilton's equations of motion, this condition becomes \cite{Arnold}
\begin{align}
  d &\omega_{\rm PC}  \wedge  \omega_{\rm PC}  \nonumber \\
   & \eqEOM   \left( H  -  p_i \frac{\partial H}{\partial p_i} \right)  dp_j  \wedge  dx^j  \wedge  dt  =  0.
\label{condExactInt}
\end{align}
Here the symbol $\eqEOM$ means the equality modulo the equation of motion (EOM) and used in this section in order to avoid possible confusions.
A simple proof of \eqref{condExactInt} is given in Appendix \ref{theproof}.
Consequently, the Pfaffian system \eqref{Pfaffsys} is completely integrable if the condition
\begin{align}
 H  \eqEOM p_i \, \frac{\partial H}{\partial p_i},
  \label{cond-exactInt}
\end{align}
is satisfied.
From Euler's theorem on homogeneous functions, this condition means that the Hamiltonian $H$ is a homogeneous function of first order in the variables $p_i$,
i.e.,
\begin{align}
 H(x, \lambda \, p, t) &= H(x, \lambda \, p_1,  \lambda \ p_2, \ldots, \lambda \, p_N, t) \nonumber \\
 &= \lambda H(x, p, t),
\end{align}
for a real $\lambda > 0$. 
With the help of the Hamilton equations $d x^i / dt = \partial H / \partial p_i$,
the associated Lagrangian 
\begin{align}
 L\left( x, \frac{d x}{dt}, t \right) := p_i \, \frac{d x^i}{dt} - H( x, p, t),
\end{align}
is null, i.e., $L \eqEOM 0$ \footnote{More generally, Lagrangian is determined up to addition of a function containing the total derivative with respect to time, i.e., for a given function $f(x^i, t)$, the Euler-Lagrange equations for $L(x, d x / dt, t) + d f(x, t) / dt$ and those for $L(x, dx/ dt, t)$ are same.}.
This does not mean the Lagrangian $L$ is algebraically null. It means that in the sense of the equality modulo EOM, i.e., for a solution of the Euler-Langrange equations (or  Hamilton's equations of motion), the value of this $L$ becomes zero.

It is known that the action $S_a(x, t)$ satisfies the following relations \cite{Arnold},
\begin{align}
  \frac{\partial S_a(x, t)}{\partial x^i}  = p_i, \quad \frac{\partial S_a(x, t) }{\partial t} = -E(t),
  \label{actionRel}
\end{align}
where $E(t)$ is the total energy of the system at a time $t$.  It follows that
\begin{align}
  d  S_a(x, t) &=  \frac{\partial S_a(x, t) }{\partial x^i} \, d x^i +  \frac{\partial S_a(x, t) }{\partial t} \, dt \nonumber \\ 
  &= p_i dx^i - E(t) dt,
  \label{dSa}
\end{align}
It is also known that the action $S_a$ and the Lagrangian $L$ are related by
\begin{align}
  d S_a(x, t) =  L \left( x, \frac{d x}{dt}, t \right) \, dt.
\end{align}
We see that the complete integrability of the Pfaffian system \eqref{Pfaffsys} leads to
\begin{align}
 \omega_{\rm PC} = dS_a(x, t) = L \, dt = 0.
  \label{sol_Pfaff}
\end{align}
From this relation and by using \eqref{dSa}, we have
\begin{align}
  d S_a( x, t) =  \left( p_i \, \frac{dx^i}{dt} - E(t) \right)  dt = 0.
\end{align}
Consequently it follows that
\begin{align}
  E(t) = p_i \, \frac{d x^i}{dt} \eqEOM H( x(t), p(t), t),
 \label{pdqH}
\end{align}
where the last expression means the instantaneous value of the Hamiltonian for a solution of the associated Hamilton's equations of motion. 
At this point we emphasize that one needs an explicit expression of $H(x, p, t)$ as a function of the canonical variables $(x, p)$ and the parameter $t$, not the value of $H$, in order to describe the associated Hamilton dynamics.

An example of the explicit expressions of Hamiltonians which are homogeneous of first order in the variables $p_i$ is
\begin{align}
   H_h( x, p, t) := \xi(t) \, c \, \sqrt{ g^{jk}(x) \, p_j p_k},
  \label{Hh}
\end{align}
where $c$ is the speed of light in vacuum, $\xi(t)$ is a dimensionless factor depending on $t$, and $g^{jk}(x)$ is the inverse of a given metric $g_{jk}(x)$ on a smooth manifold $\mathcal{M}$, i.e., $g^{jk}( x) \, g_{k\ell}(x) = \delta^j_{\ell}$.
Note that since
\begin{align}
  p_i \frac{\partial H_h}{\partial p_i} = \frac{\xi(t)\, c \, g^{i j}(x) \, p_i p_j,}{\sqrt{g^{k \ell}(x) \, p_k p_{\ell}}} = H_h,  
\end{align}
the Hamiltonian \eqref{Hh} satisfies the condition \eqref{cond-exactInt} for any $\xi(t)$. In other words, the proportional factor $\xi(t)$
is not determined by the condition \eqref{cond-exactInt} only.

Recall that the energy-momentum relation (or on shell relation) of a particle with a rest mass $m$ is
\begin{align}
  E_{\rm rel}(t) &= \sqrt{c^2  p^2(t) + m^2 c^4} \nonumber \\ 
  &=   c \, \sqrt{p^2(t)} \, \sqrt{1 + \frac{m^2 c^2 }{  p^2(t)}},
  \label{relE}
\end{align}
where $p^2(t) = g^{ij}(x)  p_i p_j $.
By substituting the well known relation
\begin{align}
  p^2(t) = \gamma^2 \, m^2 v^2(t), \; \textrm{with} \; \frac{1}{\gamma} := \sqrt{1-\frac{  v^2(t)}{c^2} },
\end{align}
in the theory of relativity, into \eqref{relE}, we see that
\begin{align}
 E_{\rm rel}(t) &=\sqrt{1 + \frac{1}{\gamma^2} \, \frac{c^2 }{ v^2(t) }}   \, c \, \sqrt{ p^2(t) }  \nonumber \\
 &=  \frac{c^2}{v(t) } \,    \sqrt{g^{ij}(x) p_i p_j} =: H_{\rm rel}(x, p, t).
 \label{Hrel}
\end{align}
Comparing this with \eqref{Hh},
the energy $E(t)$ of the Hamiltonian \eqref{Hh} can be considered as the time-dependent relativistic energy of  an accelerated (or decelerated) particle whose speed is $v(t)  = c /  \xi(t) $.  From the perspective of geometric optics, the factor $\xi(t) = c / v(t)$ can be considered as the refractive index $n$
of an optical medium.
In the point-particle viewpoint \cite{DW05}, the refractive index is expressed as
\begin{align}
   n = \frac{ c \, p }{E_{\rm ph}},
\end{align}
where $p$ and $E_{\rm ph}$ are the photon momentum and energy, respectively. It is worth mentioning  that $n = c / v_p$ where $v_p$ is a particle (photon) velocity, not a phase velocity
in the wave theory.

Note that in the natural unit $c=1$, from the above homogeneous Hamiltonian \eqref{Hrel} we can construct the following null Hamiltonian
\begin{align}
   0 &=  v(t)  \Big(  H_{\rm rel}(x, p, t) - E_{\rm rel}(t)  \Big) \nonumber \\ 
   &= \sqrt{ g^{ij}(x) \,   p_i p_j}  - p(t),
   \label{extH}
\end{align}
where $p(t) := \sqrt{ g^{ij}(x) \,   p_i p_j}$ is an instantaneous value of the momentum.
The expression in \eqref{extH} is the same form of the Hamiltonian \eqref{OurH} describing the gradient-flows in IG if we set $x^i = \theta^i, p_i = \eta_i, g^{ij}(x) = g^{ij}(\theta)$, and $p(t) = \eta(\theta(t))$.

\section{The motions of a light-like particle in a pseudo Riemann space}
\label{Null}

In general relativity \cite{Blau}, it is assumed that light propagates along a null geodesic in a pseudo-Riemann space.
An eikonal equation is assumed to be satisfied and such a light propagation follows Fermat's principle \cite{F13} and is well described in geometric optics \cite{Holm}. 
The Arnowitt, Deser, Misner (ADM) formalism \cite{ADM} is a Hamiltonian formulation of general relativity.
Caveny et al. \cite{CAM03} developed the method for tracking black hole event horizon. Their method is based on the hyperbolic eikonal equation
and it provides the Hamilton equations of motion for a null (or light-like) geodesic motion in a curved space described by a pseudo-Riemann metric.
Belayev \cite{B12} considered the variation of the energy for a light-like (null) particle in the pseudo-Riemann spacetime.
We here first review their method according to Ref.~\cite{CAM03}
and then we apply their method to the gradient-flow equations in IG by taking into account of the role of a conformal factor.

Let us consider the following form of a stationary metric
\begin{align}
  G_{\mu \nu}(x) &d x^{\mu}dx^{\nu} = - \alpha^2 (dx^0)^2 \nonumber \\ 
    & + \gamma_{ij} ( dx^i + \beta^i dx^0)(dx^j + \beta^j dx^0),
    \label{G}
\end{align}
where we assume $\alpha$ and $\beta^i$ are some functions of the space coordinate $x^i$ only and $\gamma_{ij}$ are the components of a space metric.
This form is known as $3\!+\!1$ decomposition (three~space- and one~time-coordinates) or ADM-decomposition and
\begin{align}
G_{\mu \nu}(x) &=
\left( \begin{array}{cc}
   G_{00} & G_{0j} \\ 
   G_{i 0}  & G_{ij}
\end{array}
\right) \nonumber \\
&= \left( \begin{array}{cc}
   -\alpha^2 + \gamma_{ij}\beta^i \beta^j & \gamma_{ij} \beta^i \\ 
   \gamma_{ij} \beta^j   & \gamma_{ij}
\end{array}
\right)
,
\end{align}
where $G_{\mu \nu}(x) $ are the components of the metric $G$.
The associated Lagrangian for the affine parameter $\tau$ is
\begin{align}
  L \left(x,  \frac{d x}{ d \tau} \right) = \frac{1}{2} G_{\mu \nu}(x) \frac{d x^{\mu}}{d \tau} \frac{d x^{\nu}}{d \tau}.
\end{align}
Since the canonical momenta are
\begin{align}
  p_{\mu} := \frac{\partial L}{\partial (d x^{\mu} / d \tau)} = G_{\mu \nu}(x) \frac{d x^{\nu}}{ d \tau},
\end{align}
it follows that
\begin{align}
\omega^2 :=  G^{\mu \nu}(x) p_{\mu} p_{\nu} = G_{\mu \nu}(x) \frac{d x^{\mu}}{d \tau} \frac{d x^{\nu}}{d \tau}.
\label{omega2}
\end{align}
Here the components of the inverse metric $G^{-1}$ are
\begin{align}
G^{\mu \nu}(x) =
\left( \begin{array}{cc}
  - \frac{1}{\alpha^2} & \frac{\beta^i}{\alpha^2} \\[1ex] 
   \frac{\beta^j}{\alpha^2}    & \gamma^{ij} - \frac{\beta^i \beta^j}{\alpha^2} 
\end{array}
\right)
,
\end{align}
where $\gamma^{ij}$ are the inverse matrix elements for $\gamma_{ij}$.
Since $G^{\mu \nu}(x)$ are independent of the affine parameter $\tau$, the value $\omega^2$ in \eqref{omega2} is a constant.
For $\omega^2 <0, \omega^2=0, \omega^2>0$, the metric is said to be time-like, null (or light-like), or space-like, respectively.
Since we would like to consider the motions of a light-like particle, we focus on the null case
\begin{align}
 &0 = G^{\mu \nu}(x) p_{\mu} p_{\nu}  \nonumber \\
 & \!=\! \frac{1}{\alpha^2} \! \left\{  -p_0^2 \!+\! 2 \beta^i p_i  p_0 
  \!+\! \left(\alpha^2 \gamma^{ij} \!-\! \beta^i \beta^j \right) p_i p_j \right\} \!,
 \label{null-cond}
 \end{align}
 where $1/\alpha^2$ is the conformal factor.
Solving \eqref{null-cond} for $p_0$ leads to
\begin{align}
 p_0 = \beta^i p_i \pm \sqrt{ \alpha^2 \gamma^{i j} p_i p_j }.
\end{align}
We then express \eqref{null-cond} as
\begin{align}
  \frac{1}{\alpha^2} \left\{ H^+ (x, p) H^-(x, p) \right\} = 0,
\end{align}
where 
\begin{align}
 H^{\pm}(x, p) := p_0 -\beta^i p_i \pm \sqrt{ \alpha^2 \gamma^{i j}  p_i p_j }.
 \end{align}
 Thus the null Hamiltonian for a light-like particle is either $H^{+}(x, p) =0$ or $H^{-}(x, p) =0$.
 Then the corresponding Hamilton equations of motion are
\begin{align*}
  \frac{dx^0}{d \tau} = H^{\mp} \frac{\partial H^{\pm} }{\partial p_0 } = H^{\mp}, \quad \frac{d x^i}{d \tau} = H^{\mp} \frac{\partial H^{\pm} }{\partial p_i }, \\
  \frac{d p_0}{d \tau} = -H^{\mp} \frac{\partial H^{\pm} }{\partial x^0} = 0, \quad \frac{d p^i}{d \tau} = -H^{\mp} \frac{\partial H^{\pm} }{\partial x^i },
\end{align*}
where the upper- (lower-) case in the superscript refers to the choice $H^+ = 0$ ( $H^- = 0$).
By using these relations we can eliminate the affine parameter $\tau$ and obtain the equations of motion with respect to $x^0$ as follows.
\begin{subequations}
\begin{empheq}[left = {\empheqlbrace \,}, right = {}]{align}
  \frac{d x^i}{dx^0} &= \frac{ \frac{d x^i }{ d \tau}}{\frac{dx^0 }{ d \tau}} = \frac{\partial H^{\pm} }{\partial p_i }  \nonumber \\
  &= \frac{\partial}{\partial p_i } \left\{ -\beta^j p_j \pm \sqrt{ \alpha^2 \gamma^{j k} p_j p_k }  \right\}, 
  \\
  \frac{d p_i}{dx^0} &= \frac{\frac{d p_i }{ d \tau}}{\frac{dx^0 }{ d \tau} } = -\frac{\partial H^{\pm} }{\partial x^i } \nonumber \\
   &= -\frac{\partial}{\partial x^i } \left\{ -\beta^j p_j \pm \sqrt{ \alpha^2 \gamma^{j k} p_j p_k }  \right\}.
\end{empheq}
\end{subequations}
Consequently the expression in the curly brackets acts as the Hamilton function
\begin{align}
\mathcal{H}^{\pm}(x, p) := -\beta^j p_j \pm \sqrt{ \alpha^2 \gamma^{j k} p_j p_k },
\label{associatedH}
\end{align}
which describes the motions of a null particle with respect to the parameter $x^0$.
 
\subsection{Relation to the Randers-Finsler Lagrangian}
Here we explain the relationship between the Hamiltonian \eqref{associatedH} and  the RF Lagrangian \eqref{RFLag}.
It is well known that the Legendre transformation maps a Hamiltonian to a Lagrangian.
However, since $\mathcal{H}^{\pm}(x, p)$ is homogeneous of first order in momenta $p$, the associated Lagrangian would vanish.
A useful method \cite{GHWW09} is introduce a Hamiltonian, say $\mathcal{G}(x, p)$, as
\begin{align}
   \mathcal{G}(x, p)  = \frac{1}{2} \left( \mathcal{H}^{+} \right)^2,
\end{align}
which is homogeneous of second order in momenta $p$.  Then the Legendre transformation of $\mathcal{G}$ leads to the Lagrangian
\begin{align}
  L = \frac{1}{2} \mathcal{F}^2, 
\end{align}
where $\mathcal{F}=\mathcal{F}(x, v)$ is a homogeneous function of first order in velocities $v^i$, i.e., $\mathcal{F} = v^i  \partial \mathcal{F} / \partial v^i$.
Since $p_i = \partial L / \partial v^i$ and
\begin{align}
  v^i p_i = v^i \frac{\partial L}{\partial v^i} = \mathcal{F} v^i \frac{\partial \mathcal{F}}{\partial v_i} = \mathcal{F}^2, 
\end{align}
one readily find that
\begin{align}
  \mathcal{G} = p_i v^i - L = \mathcal{F}^2 - \frac{1}{2} \mathcal{F}^2 = L.
 \end{align}
As a result we see that $\mathcal{H}^+(x, p(v))  = \mathcal{F}(x, v(p))$.
 
Next the expression of $\mathcal{F}(x, v)$ is obtained as follows.
From Hamilton's equations of motion for $\mathcal{G}$, we have
\begin{align}
 v^i &:= \frac{d x^i}{d x^0} = \frac{\partial \mathcal{G}}{\partial p_i} =  \mathcal{H}^{+} \, \frac{\partial \mathcal{H}^{+}}{\partial p_i}
 =  \mathcal{H}^{+} \, ( -\beta^i + \nu^i), \nonumber \\
 &\textrm{ with } \nu^i = \frac{ \tilde{\gamma}^{i j} p_j}{\sqrt{ \tilde{ \gamma}^{k \ell} p_k p_{\ell} }},
\end{align}
where $ \tilde{\gamma}^{i j} := \alpha^2 \gamma^{ij}$.
Since $\tilde{\gamma}_{ij} \nu^i \nu^j=1$ and $\nu^i = (v^i / \mathcal{H}^+) + \beta^i = (v^i / \mathcal{F}) + \beta^i $, we have
\begin{align}
  1 = \tilde{\gamma}_{ij} \left( \frac{v^i }{ \mathcal{F}} + \beta^i \right)  \left( \frac{v^j }{ \mathcal{F}} + \beta^j \right).
\end{align}
Solving for $\mathcal{F}$ and introduce the transformed metric $a_{ij}$ and $b_i$ by
\begin{alignat}{2}
 a_{ij} &= \frac{ \xi  \tilde{\gamma}_{ij} + \beta_i \beta_j}{\xi^2}, &\quad b_i &= \frac{\beta_i}{\xi}, \nonumber \\
  \xi &:= 1 - \tilde{\gamma}_{ij} \beta^i \beta^j,  &\quad \beta_i &=  \tilde{\gamma}_{ij} \beta^j,
\end{alignat}
we obtain
\begin{align}
   \mathcal{F}(x, v ) = \sqrt{ a_{ij}(x)  v^i v^j} + b_i(x) v^i,
\end{align}
which is the RF Lagrangian obtained from a Randers function \cite{Randers}.
Note that by setting $x = \theta, a_{ij} = g_{ij}(\theta) / \chi^2(\theta), b_i = A_i(\theta) / \chi^2(\theta)$ and $v^i = d \theta^i / dt$,
this $\mathcal{F}$ becomes the RF Lagrangian \eqref{RFLag}.

\subsection{Applications to the gradient-flow equations}

Firstly, we consider the gradient-flow equations \eqref{theta-gradEq} for the $\theta$-potential function $\Psi(\theta)$ in the case
of $\theta_{\rm r}^i = \eta_i^{\rm r} = 0$.
Taking the derivative of $\Psi(\theta)$ with respect to $t$ and using \eqref{theta-eta}, we have
\begin{align}
 \frac{d \Psi(\theta)}{dt} = \frac{\partial \Psi(\theta)}{\partial \theta^i} \frac{d \theta^i}{dt} = \eta_i \frac{d \theta^i}{dt} 
 = g^{ij}(\theta) \eta_i \eta_j,
\end{align}
where in the last step we used  \eqref{theta-gradEq} with $\eta^{\rm r}_j = 0$.
Now we introduce the quantity $\eta^2(\theta)$ defined in \eqref{2ndH4theta}, and rewrite the above relation as
\begin{align}
 0 =    \eta^2(\theta) \Big( \tilde{g}^{ij}(\theta) \eta_i(\theta) \eta_j(\theta) - 1 \Big), 
 \label{null-rel}
\end{align}
where we introduced the conformal metric
\begin{align}
  \tilde{g}^{ij}(\theta) := \frac{1}{ \eta^2(\theta)} \,  g^{ij}(\theta).
\end{align}
We can regard the null relation \eqref{null-rel} as \eqref{null-cond} by setting $p_i = \eta_i, p_0 = -1, \alpha^2 =  1/ \eta^{2}(\theta), \beta^i=0$, and $\gamma^{ij} = g^{ij}(\theta)$. 
Then the corresponding Hamiltonian $\mathcal{H}^{+}$ in \eqref{associatedH} becomes
\begin{align}
\mathcal{H}^{+}( \theta, \eta) =  \sqrt{\tilde{g}^{ij}(\theta) \, \eta_i \eta_j} = \sqrt{\frac{g^{ij}(\theta)}{\eta^2(\theta)} \,  \eta_i \eta_j} ,
\label{IGH}
\end{align}
which describes the gradient-flows in IG as a light-like particle motion in the pseudo-Riemann space with 
$G^{\mu \nu}(x) =\eta^2(\theta) \, {\rm diag.}(-1,  \tilde{g}^{ij}(\theta))$.
The conformal factor $1/ \alpha^2 =  \eta^{2}(\theta)$ does not affect the light-like geodesic but change their parametrization \cite{Cariglia15}, which
is explained in Appendix \ref{conformal}. 
The corresponding Hamilton's equations of motion are
    \begin{empheq}[left = {\empheqlbrace \,}, right = {}]{align}
\label{EOM4Hp}
  &\frac{d \theta^i}{d x^0} = \frac{\partial \mathcal{H}^+ }{\partial \eta_i } 
  = \frac{1}{ \eta^2(\theta)} \, g^{ij}(\theta) \eta_j,  \nonumber
  \\
 & \frac{d \eta_i}{d x^0} = -\frac{\partial \mathcal{H}^+ }{\partial \theta^i }  \nonumber \\
  &= \frac{1}{ 2 \eta^2(\theta)}  \left( - \frac{\partial g^{jk}(\theta) }{\partial \theta^i } \eta_j \eta_k + \frac{\partial \eta^2(\theta) }{\partial \theta^i } \right).
\end{empheq}
We can change the parametrization from $x^0$ to $t$ according to
\begin{align}
  dx^0 = \eta^2(\theta) dt.
  \label{dt}
\end{align}
This maps  the equations \eqref{EOM4Hp} to those for the Hamiltonian \eqref{H4theta}.
In this way we have rederived the Hamiltonian \eqref{H4theta} describing the gradient-flows in IG. 
The other Hamiltonian \eqref{H4eta} can be obtained in a similar way.

Secondly, we consider the RF deformed gradient-flow equations \eqref{RFtheta-gradEq}.
Arranging the relation \eqref{rel_chi2} leads to
\begin{align}
  0 = g^{ij}(\theta) \eta_i \eta_j  \! - \!  2 A^i(\theta) \eta_i  \! - \!  (\chi^2(\theta) \! - \! A^2(\theta) ),
  \label{4RFdeform}
\end{align}
where we used $A^i (\theta) = g^{ij}(\theta) A_j(\theta)$ and $A^2(\theta) := g^{ij}(\theta) A_i(\theta) A_j(\theta)$.
By setting
\begin{align}
   g^{ij}(\theta) =  \xi \chi^2(\theta) ( \tilde{\gamma}^{ij} - \beta^i \beta^j), \; \; \; \beta^i = \frac{A^i(\theta)}{ \xi \chi^2(\theta)},
\end{align}
and using $\xi = 1 - A^2(\theta) / \chi^2(\theta)$, the above relation \eqref{4RFdeform} is rewritten as
\begin{align}
  0 = \xi \chi^2(\theta) \left[ (\tilde{\gamma}^{ij} - \beta^i \beta^j) \eta_i \eta_j - 2 \beta^i \eta_i - 1 \right].
\end{align}
We again regard this null relation as \eqref{null-cond} by setting $p_i = \eta_i, p_0 = -1$ and $1/ \alpha^2 = \xi  \chi^2(\theta)$.
Then the corresponding Hamiltonian $\mathcal{H}^{+}$ in \eqref{associatedH} becomes
\begin{align}
\mathcal{H}^{+} &( \theta, \eta) = - \frac{A^i(\theta)}{\xi \chi^2(\theta)}  \eta_i  \nonumber \\
&+  \sqrt{ \left(  \frac{g^{ij}(\theta)}{\xi \chi^2(\theta)}  + \frac{ A^i(\theta)}{\xi \chi^2(\theta)}  \frac{ A^j(\theta)}{\xi \chi^2(\theta)}   \right)\, \eta_i \eta_j}.
\label{RF_H}
\end{align}

As we mentioned in a few lines after \eqref{RFtheta-gradEq}, the gradient-flow equations \eqref{theta-gradEq} correspond to the RF deformed equations \eqref{RFtheta-gradEq}
when $A_j(\theta) = \eta^{\rm r}_j$. Consequently, the corresponding Hamiltonian for \eqref{theta-gradEq} is obtained by replacing $A^i(\theta) = g^{ik}(\theta) A_k(\theta)$ with $g^{ik}(\theta) \eta^{\rm r}_k$, and $\xi = 1-A^2(\theta)/\chi^2(\theta)$ with $\xi = 1-(\eta_r)^2 / \chi^2(\theta)$ in \eqref{RF_H}.

\subsection{Gaussian model}
\label{Gaussian}
As a concrete example of the gradient-flows, we here consider the Gaussian, or Normal $N(\mu, \sigma^2)$, pdf which is given by
\begin{align}
 p_{\rm G}(x; \mu, \sigma) = \frac{1}{\sqrt{2 \pi \, \sigma^2}} \,
     \exp \left[ -\frac{(x-\mu)^2}{2 \sigma^2} \right].
\end{align}
Here $\mu$ denotes the mean and $\sigma^2$ is the variance. 
It is known that the natural $\theta$-coordinates and $\eta$-coordinates \cite{Amari} are \footnote{Do not confuse the superscript in $\theta$ variables  with exponents.}
\begin{align}
 \theta^1 &= \frac{\mu}{\sigma^2}, \quad \theta^2 = -\frac{1}{2 \sigma^2}, \nonumber \\
 \eta_1 &= \mu, \quad \eta_2 = \mu^2 + \sigma^2.
\end{align}
The components $g_{ij}(\eta)$ of the metric tensor $g(\eta)$ are 
\begin{align}
g_{ij}(\eta) &=
 2 \big( \eta_2 - (\eta_1)^2 \big)
\left( \begin{array}{cc}
   \frac{1}{2} & \eta_1 \\[1ex] 
    \eta_1  &  ( \eta_1)^2 + \eta_2
\end{array}
\right) \nonumber \\
&=
2 \sigma^2
\left( \begin{array}{cc}
   \frac{1}{2} & \mu  \\[1ex] 
    \mu  &  2 \mu^2 + \sigma^2
\end{array}
\right)
.
\end{align}
The linear differential equation \eqref{linearDE} of the gradient-flow equations \eqref{eta-gradEq} are
\begin{subequations}
\label{GradFeqs}
\begin{empheq}[left=\empheqlbrace]{align}
   \frac{d}{d t} \theta^1 &= \frac{1}{\sigma^2} \frac{d \mu }{d t} -  \frac{2 \mu}{\sigma^3} \frac{d \sigma}{dt} = -\frac{\mu}{\sigma^2}  + \frac{\mu_{\rm r}}{\sigma_{\rm r}^2}, \\
\frac{d}{d t} \theta^2 &=  \frac{1}{\sigma^3} \frac{d \sigma}{d t} =   \frac{1}{2 \sigma^2} - \frac{1}{2 \sigma_{\rm r}^2},
\end{empheq}
\end{subequations}
where the set of $\mu_{\rm r}$ and $\sigma_{\rm r}$ specify the reference state, whose $\theta$-coordinates are $\theta_{\rm r}^1 = \mu_{\rm r} / \sigma_{\rm r}^2$
and $\theta_{\rm r}^2 = - 1 /( 2  \sigma_{\rm r}^2)$.
From \eqref{GradFeqs}, we obtain the differential equations for $\mu(t)$ and $\sigma(t)$ as
\begin{subequations}
\label{GradF_mu_sigma}
\begin{empheq}[left=\empheqlbrace]{align}
   \frac{d}{d t} \mu(t) &= \frac{\sigma^2}{\sigma_{\rm r}^2} ( \mu_{\rm r} -  \mu(t)), \\
\frac{d}{d t} \sigma(t) &=  \frac{1}{2} \left(   \sigma(t)  - \frac{\sigma^3(t)}{ \sigma_{\rm r}^2} \right),
\end{empheq}
\end{subequations}
and the solutions are
\begin{subequations}
\begin{empheq}[left=\empheqlbrace]{align}
   \mu(t) &= \frac{\mu_0 \sigma_{\rm r}^2 + \mu_{\rm r} \sigma_0^2 (\exp(t)-1)}{\sigma_{\rm r}^2 + \sigma_0^2(\exp(t)-1)}, \\
\sigma(t) &= \frac{\sigma_{\rm r} \sigma_0 \exp \left( \frac{t}{2} \right)}{ \sqrt{\sigma_{\rm r}^2 + \sigma_0^2(\exp(t)-1)}},
\end{empheq}
\end{subequations}
where $\mu_0$ and $\sigma_0$ are the initial values, i.e., $\mu(0) = \mu_0$ and $\sigma(0) = \sigma_0$.  Note that
$\mu_{\rm r}$ and $\sigma_{\rm r}$ are the final values, i.e., $\lim_{t \to \infty} \mu(t) = \mu_{\rm r}$ and $\lim_{t \to \infty} \sigma(t) =\sigma_{\rm r}$.

\begin{figure}[h]
\centering
 \includegraphics[width=70mm]{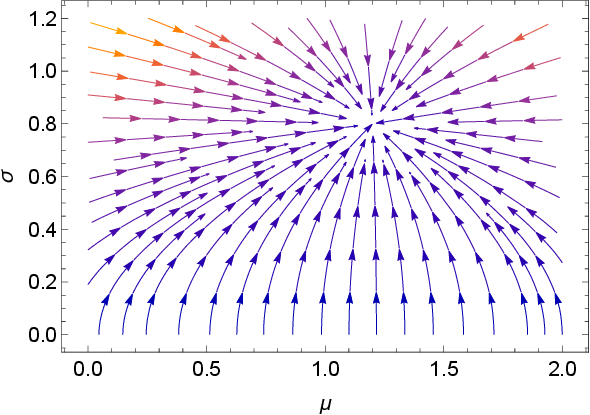}
\caption{The gradient-flows of the equations \eqref{GradF_mu_sigma} with $\mu_{\rm r} = 1.2, \sigma_{\rm r}=0.8$. }
\label{Fig1}
\end{figure}
Fig. \ref{Fig1} shows the gradient-flows of \eqref{GradF_mu_sigma} in $(\mu, \sigma)$-space. The reference state is specified by $\mu_{\rm r} = 1.2, \sigma_{\rm r}=0.8$,
at which $\eta= \eta_{\rm r}$ and the divergence $D(\eta, \eta_{\rm r})$ vanishes.


From \eqref{GradF_mu_sigma}, we  also obtain the explicit relations $dt$, $d\mu$ and $d\sigma$ along these gradient-flows as 
\begin{align}
    d t =\frac{\sigma_{\rm r}^2}{\sigma^2} \frac{ d \mu}{\mu_{\rm r}- \mu} =  \frac{ 2 d \sigma}{\sigma \left( 1 - \frac{\sigma^2}{\sigma_{\rm r}^2} \right)}.
\end{align}

\section{Conclusions and perspectives}
\label{conclusions}

We have related the motions of a light-like particle in a curved space to
the gradient-flows in IG. Based on the point-particle viewpoint, we have rederived the Hamiltonians  \eqref{OurH}  in our previous works \cite{WSM23, W23}.
In addition, it is shown that the complete integrability of Pfaffian systems for the Poincar\'e-Cartan one-form supports this type of Hamiltonian \eqref{OurH}.

As mentioned in Introduction, our studies on the gradient-flow equations in IG have some relations to different fields such as analytical mechanics, geometric optics, thermodynamics, general relativity, cosmology and so on.
For examples, in regard to thermodynamics, a linear constitutive relation for non-equilibrium thermodynamics can be written in the form
\begin{align}
 \frac{d X_i}{dt} =  L_{ij}(X)   \frac{\partial }{\partial X_j} S(X),
 \label{OnsagerEq}
\end{align}
where $X_i$ denotes an extensive variable, $S(X)$ is the thermodynamic entropy, and $L_{ij}(X)$ are the components of Onsager's phenomenological matrix.
Note that \eqref{OnsagerEq} can be regarded as the gradient-flow equations \eqref{eta-gradEq} if we make the correspondence that $X_i \leftrightarrow \eta_i, L_{ij} \leftrightarrow g_{ij}(\eta)$, and $S(X) \leftrightarrow  - \Psi^{\star}(\eta)$.
In this correspondence, famous Onsager's reciprocal relations $L_{ij} = L_{jk}$
can be understood in IG as the symmetry of the metric $g_{ij}(\eta) = \partial \eta_i / \partial \theta^j = \partial \eta_j / \partial \theta^i = g_{ji}(\eta) $, which is 
due to integrability. Recently, Katagiri \cite{Katagiri} extended the constitutive relations of Onsager's non-equilibrium thermodynamics by considering
a thermodynamic force as a gauge fixing. It is noticed that some relations in Ref. \cite{Katagiri} can be regarded as the RF deformed gradient-flow equations in IG.

In regard to general relativity and cosmology, Ref. \cite{CW24} described the dynamical evolutions of flat metrics for Kerr and Reissner-Nordstr\"om black holes, and Ref. \cite{W23} shows that the significance of Weyl integrable geometry in IG.
Moreover Gibbons et al. \cite{GHWW09}  showed the triality among the Zermelo navigation problem, the geodesic flow on a RF function, and optical metric of one dimension higher stationary spacetime. The Zermelo/Randers/spacetime triality allows us to translate one of the three viewpoints to any of the other two viewpoints, resulting in significant simplifications or complications. Since the stationary metric \eqref{G} is equivalent to the Zermelo form Eq. (31) in Ref. \cite{GHWW09} as shown in Appendix \ref{Zform}, it will be interesting to study the gradient-flows in IG from these viewpoints.

Furthermore, in regard to the approaches to the gradient-flows in IG based on analytical mechanics, Pistone et al.  \cite{CMP22,Pistone} have been developed their Lagrangian and Hamiltonian
formalism. It is intriguing to study the relation between their method and this work.
Finally we believe that it is worthwhile to further explore the information geometric studies on the gradient-flow equations from some different perspectives in the fields of physics.

\bmhead{Acknowledgements}

The first named author (T.W.) was supported by Japan Society for the Promotion of Science (JSPS) Grants-in-Aid for Scientific Research (KAKENHI) Grant Number JP22K03431.
We thank the anonymous referees for their valuable comments.

\subsection*{Declarations}
\subsection*{Conflict of interest}
 The authors have no relevant financial
or non-financial interests to disclose. 

\subsection*{Author contribution}
Tatsuaki Wada: Conceptualization, Methodology, Validation, Writing--original draft, review and editing.
Antonio Maria Scarfone: Conceptualization, Validation, Writing--review and editing.

\subsection*{Data availability statement}
No associated data.

\begin{appendices}

\section{The proof of \eqref{condExactInt}}
\label{theproof}
In the extended phase space of $(x, p, t) \in T^{\star} \mathcal{M} \times \mathbb{R}$,
the canonical equation of motion can be expressed as
\begin{align}
   \iota(X) \, d\omega_{\rm PC} = 0,
\end{align}
where $X$ is a vector field in the extended phase space and $\iota(X)$ denote the interior product.
More concretely we have
\begin{align}
   -\iota \! \left( \frac{\partial}{\partial x^i} \right) \! d\omega_{\rm PC} &= dp_i + \frac{\partial H}{\partial x^i} dt = 0, \nonumber \\
   \iota \! \left( \frac{\partial}{\partial p_i} \right) \! d\omega_{\rm PC} &= dx^i - \frac{\partial H}{\partial p_i} dt = 0,
   \label{iprod}
\end{align}
as Hamilton's equations of motion.

Now we consider the complete integrability of the Pfaffian equation \eqref{Pfaffsys}.
Recall that a differential form $\alpha$ is \textit{closed} if its exterior derivative is zero ($d \alpha=0$),
and is \textit{exact} if it is the exterior deriative of another differential form ($\alpha = d \beta$).
From \eqref{PCform} we have
\begin{align}
 d\omega_{\rm PC} = dp_j \wedge dx^j - dH(x, p, t) \wedge dt.
 \label{domega}
\end{align}
By using \eqref{iprod} we have
\begin{align}
 d& H \wedge dt = \left( dx^i \frac{\partial H}{\partial x^i} +   dp_i \frac{\partial H}{\partial p_i} + dt \frac{\partial H}{\partial t} \right) \wedge dt  \nonumber \\
 &=
dx^i \wedge \frac{\partial H}{\partial x^i} dt +  dp_i \wedge \frac{\partial H}{\partial p_i} dt \nonumber \\
&\eqEOM dx^i \wedge (-dp_i) \!+\! dp_i \wedge dx^i \!=\! 2 dp_i \wedge dx^i.
\end{align}
then
\eqref{domega} becomes $d\omega_{\rm PC} \eqEOM -dp_j \wedge dx^j$.
It follows that
\begin{align}
  d \omega_{\rm PC} &\wedge \omega_{\rm PC} 
 \eqEOM H dp_i \wedge dx^i \wedge dt - p_j dx^j \wedge dp_i \wedge dx^i \nonumber \\
&\eqEOM H dp_i \wedge dx^i \wedge dt - p_j \frac{\partial H}{\partial p_j} dt \wedge dp_i \wedge dx^i \nonumber \\
&= \left(H - p_j \frac{\partial H}{\partial p_j} \right) dp_i \wedge dx^i \wedge dt,
\end{align}
where we used \eqref{iprod}.

\section{Conformal rescaling as reparametrization}
\label{conformal}

Here we briefly explain the relation between the conformal rescaling and reparametrization
of null geodesics \cite{Cariglia15}.
For a given indefinite metric $g_{\mu \nu}(x)$, consider the null geodesic Hamiltonian
\begin{align}
  \mathcal{H} = \frac{1}{2}  g^{\mu \nu}(x) p_{\mu} p_{\nu} = 0.
\end{align}
Its null geodesics are the solutions of Hamilton's equations of motion,
\begin{empheq}[left = {\empheqlbrace \,}, right = {}]{align}
  \frac{d x^{\mu}}{d t} &= \frac{\partial \mathcal{H}}{\partial p_{\mu}} = g^{\mu \nu}(x) p_{\nu},\\
  \frac{d p_{\mu}}{d t} &= -\frac{\partial \mathcal{H}}{\partial x^{\mu}} = -\frac{1}{2} \frac{\partial g^{\nu \rho}(x)}{\partial x^{\mu}} p_{\nu} p_{\rho},
\end{empheq}
where $t$ is the evolution parameter.
Now consider the rescaled metric
\begin{align}
 \tilde{g}_{\mu \nu}(x) = \Omega^2(x) g_{\mu \nu}(x),
 \label{rescaling}
 \end{align}
 and the associated new Hamiltonian
\begin{align}
  \tilde{ \mathcal{H}} = \frac{ g^{\mu \nu}(x)}{2 \Omega^{2}(x)} \, p_{\mu} p_{\nu} = 0,
\end{align}
where $\Omega^2(x) > 0$ is a scaling factor. The new equations of motion are
\begin{empheq}[left = {\empheqlbrace \,}, right = {}]{align*}
 \frac{d x^{\mu}}{d \tilde{t}} &= \frac{\partial \tilde{\mathcal{H}}}{\partial p_{\mu}} = \frac{1}{ \Omega^{2}(x)} g^{\mu \nu}(x) p_{\nu} =  \frac{1}{ \Omega^{2}(x)}  \frac{d x^{\mu}}{d t},\\
 \frac{d p_{\mu}}{d \tilde{t}}  &= -\frac{\partial \tilde{\mathcal{H}} }{\partial x^{\mu}} \nonumber \\
   &= -\frac{1}{2 \Omega^{2}(x)} \frac{\partial g^{\nu \rho}(x)}{\partial x^{\mu}} p_{\nu} p_{\rho}
  -\mathcal{H} \frac{\partial }{\partial x^{\mu}}  \frac{1}{\Omega^{2}(x)}  \nonumber \\
  & =  \frac{1}{\Omega^{2}(x)}  \frac{d p_{\mu}}{d t},
\end{empheq}
where $\tilde{t}$ is the evolution parameter in the new $\tilde{\mathcal{H}}$ and we used $\mathcal{H}=0$ in the last step.
From these relations, one sees that both Hamiltonians share the same null geodesics and changing the evolution parameter according to
\begin{align}
   d \tilde{t} = \Omega^2(x) dt.
   \label{repara}
\end{align}
This maps the equations of motion for $\tilde{\mathcal{H}}$ into those fo $\mathcal{H}$.
In this way, the conformal rescaling  \eqref{rescaling} is functioning as the change of the evolution parameter \eqref{repara}.

\section{Zermelo form}
\label{Zform}
The Zermelo navigation problem is a time-optimal control problem, which aims at finding the shortest-time path under the influence
of a window vector $W^i$.
We here show the explicit relations between the stationary metric \eqref{G} and the Zermelo form:
\begin{align}
  &ds^2 = \frac{V^2}{1-h_{ij} W^i W^j}  \nonumber \\
    & \times  \left[ -dt^2  \! + \!  h_{ij} (dx^i \! - \! W^i dt)(dx^j \! - \! W^j dt) \right],
    \label{Zermelo}
\end{align}
which is Eq. (31) in Ref. \cite{GHWW09}.
By setting
\begin{align}
    V^2 &:= \alpha^2 - \gamma_{ij} \beta^i \beta^j, \quad h_{ij} := \tilde{\gamma}_{ij} = \frac{\gamma_{ij}}{\alpha^2},\nonumber \\
     W^i &:= -\beta^i, \quad dt = dx^0,
\end{align}
we have
\begin{align}
  \frac{V^2}{ 1- h_{ij}W^i W^j} = \frac{\alpha^2 - \gamma_{ij}\beta^i \beta^j}{1 - \frac{\gamma_{ij}}{\alpha^2} \beta^i \beta^j} = \alpha^2.
 \end{align}
 Then the Zermelo form \eqref{Zermelo} becomes  \eqref{G}.




\end{appendices}



\begin{thebibliography}{9}

\bibitem{Amari}
S-I. Amari,
{\it Information geometry and its applications},
{\it Appl. Math. Sci.} {\bf 194} (Springer, Tokyo, 2016)

\bibitem{WS15}
T. Wada, A.M. Scarfone, 
Information geometry on the $\kappa$-thermostatistics.
 Entropy  \textbf{17}, 1204-1217 (2015)

\bibitem{N94}
Y. Nakamura,
Gradient systems associated with probability distributions.
\textit{Japan J. Indust. Appl. Math.}
\textbf{11} 21-30 (1994)

\bibitem{FA95}
A. Fujiwara, S-I. Amari, 
Gradient systems in view of information geometry.
\textit{Physica} D \textbf{80} 317 (1995)

\bibitem{MP15}
Malag\'o L. and Pistone G.,
Natural Gradient Flow in the Mixture Geometry of a Discrete Exponential Family.
\textit{Entropy} \textbf{17} 4215--4254 (2015)

\bibitem{BN16}
N. Boumuki, T. Noda,
On gradient and Hamiltonian flows on even dimensional dually flat spaces.
\textit{Fundamental J. Math. and Math Sci. } \textbf{6} 51-66 (2016)

\bibitem{CMP22}
G. Chirco, L. Malag\'o, G. Pistone,
Lagrangian and Hamiltonian Mechanics for Probabilities on the Statistical Manifold.
arXiv:2009.09431v2



\bibitem{Pistone}
G. Pistone,
Lagrangian Function on the Finite State Space Statistical Bundle.
\textit{Entropy}, \textbf{20}, 139  (2018) \\
\url{https://doi.org/10.3390/e20020139}


\bibitem{WSM21}
T. Wada, A.M. Scarfone, H.  Matsuzoe,  
An eikonal equation approach to thermodynamics and the gradient flows in information geometry.
\textit{Physica} A, \textbf{570} 125820 (2021)


\bibitem{WSM23}
T. Wada, A.M. Scarfone, H.  Matsuzoe,    
Huygens' equations and the gradient-flow equations in information geometry.
\textit{ Int. J. Geom. Methods Mod. Phys.}, \textbf{20} 2450012 (2023)



\bibitem{CW24}
S. Chanda, T. Wada,
Mechanics of geodesics in information geometry and black hole thermodynamics.
\textit{ Int. J. Geom. Methods Mod. Phys.}, \textbf{ 21}  2450098 (2024)

\bibitem{Randers}
G. Randers,
On an asymmetrical metric in the four-space of general relativity.
\textit{Phys. Rev. } \textbf{59} 195 (1941)

\bibitem{W23}
T. Wada,  
Weyl geometric approach to the gradient-flow equations in information geometry.
\textit{J. Geom. Symmetry Phys.}, \textbf{66} 59-70 (2023)


\bibitem{Blau}
M. Blau,
\textit{Lecture Notes on General Relativity}\\
\url {http://www.blau.itp.unibe.ch/GRLecturenotes.html}



\bibitem{G18}
Y. Gu, 
Natural coordinate system in curved space-time.
\textit{J. Geom. Symmetry Phys. }\textbf{47} 51-62 (2018)


\bibitem{F13}
V.P. Frolov, 
Generalized Fermat's principle and action for light rays in a curved spacetime.
\textit{Phys. Rev. D} \textbf{88} 064039 (2013)

\bibitem{Cartan}
E. Cartan,
Chap. X, \textit{Lessons on integral invariants}
translated by D.H. Delphenich (Hermann and Co., Paris, 1922)

\bibitem{Arnold}
V.I. Arnold, 
\textit{Mathematical methods of classical mechanics}, 2nd edition,
(Springer-Verlag, New York, 1989)

\bibitem{DW05}
D. Drosdoff, A. Widom, 
Snell's law from an elementary particle viewpoint.
\textit{Am. J. Phys.}  \textbf{73} 973-975 (2005)


\bibitem{Holm}
D.D. Holm,
\textit{Fermat's principle and the geometric mechanics of ray optics}.
Summer School Lectures, Fields Institute, Toronto.  (2012)

\bibitem{ADM}
R. Arnowitt, S. Deser, C.W. Misner, 
Dynamical structure and definition of energy in general relativity.
\textit{Phys. Rev.} \textbf{116}, 1322 (1959) 

\bibitem{CAM03}
S.A. Caveny, M. Anderson, R.A. Matzner, 
Tracking black holes in numerical relativity.
\textit{Phys. Rev. D} \textbf{68} 104009 (2003)

\bibitem{B12}
W. Belayev, 
Application of Lagrange mechanics for analysis of the light-like particle motion in pseudo-Riemann space.
\textit{Int. J. Theor. Math. Phys.} \textbf{2} 10-15 (2012)

\bibitem{GHWW09}
G.W. Gibbons, C.A.R. Herdeiro, C.M. Warnick, M.C. Werner,
Stationary metrics and optical Zermelo-Randers-Finsler geometry.
\textit{Phys. Rev. D} \textbf{79} 044022 (2009)

\bibitem{Cariglia15}
M. Cariglia,
Null lifts and projective dynamics.
\textit{Annals of Physics} \textbf{362} 642-658 (2015)










\bibitem{Katagiri}
S. Katagiri, 
Non-equilibrium thermodynamics as gauge fixing.
\textit{Prog. Theor. Exp. Phys.}, \textbf{2018}, issue 9 (2018) 




\end{thebibliography}

\end{document}